%Paper: hep-th/9404004
%From: "Andrei Matytsin" <matytsin@puhep1.Princeton.EDU>
%Date: Fri, 1 Apr 94 17:03:02 -0500
%Date (revised): Mon, 25 Apr 94 16:43:46 -0400

%%%%%% TeX file, using harvmac.tex and epsf.tex %%%%%%%%%%%%%%%%%%%%%%%%%

\input harvmac.tex
\input epsf

\def\figin{\epsfcheck\figin}\def\figins{\epsfcheck\figins}
\def\epsfcheck{\ifx\epsfbox\UnDeFiNeD
\message{(NO epsf.tex, FIGURES WILL BE IGNORED)}
\gdef\figin##1{\vskip2in}\gdef\figins##1{\hskip.5in}% blank space instead
\else\message{(FIGURES WILL BE INCLUDED)}%
\gdef\figin##1{##1}\gdef\figins##1{##1}\fi}
\def\DefWarn#1{}
\def\figinsert{\goodbreak\midinsert}
\def\ifig#1#2#3{\DefWarn#1\xdef#1{fig.~\the\figno}
\writedef{#1\leftbracket fig.\noexpand~\the\figno}%
\figinsert\figin{\centerline{#3}}\medskip\centerline{\vbox{\baselineskip12pt
\advance\hsize by -1truein\noindent\footnotefont{\bf Fig.~\the\figno:} #2}}
\bigskip\endinsert\global\advance\figno by1}

\def\ee{{\rm e}}
\font\cmss=cmss10 \font\cmsss=cmss10 at 7pt
\def\IZ{\relax\ifmmode\mathchoice
{\hbox{\cmss Z\kern-.4em Z}}{\hbox{\cmss Z\kern-.4em Z}}
{\lower.9pt\hbox{\cmsss Z\kern-.4em Z}}
{\lower1.2pt\hbox{\cmsss Z\kern-.4em Z}}\else{\cmss Z\kern-.4em Z}\fi}

\lref\REFmigdal{A. Migdal, Sov. Phys. JETP {\bf 42} (1975) 413 \
(Zh. Exp. Teor. Fiz. {\bf 69} (1975) 810).}
\lref\REFrusakov{B. Rusakov, Mod. Phys. Lett. {\bf A5} (1990) 693.}
\lref\REFdouglas{M. R. Douglas and V. A. Kazakov, {\sl Large $N$ Phase
Transition in Continuum $QCD_{2}$}, preprint LPTENS-93120, RU-93-17,
hep-th 9305047.}
\lref\REFtaylor{D. Gross, Nucl. Phys. {\bf B400} (1993) 161\semi
D. Gross and W. Taylor, Nucl. Phys. {\bf B400} (1993) 181\semi
Nucl. Phys. {\bf B403} (1993) 395.}
\lref\REFminahan{J. Minahan and A. Polychronakos, {\sl Classical Solutions
for Two-Dimensional QCD on the Sphere}, preprint CERN-TH-7016/93,
UVA-HET-93-08, hep-th 9309119.}
\lref\REFDADDA{A. D'Adda, M. Caselle, L. Magnea  and S. Panzeri,
{\sl Two dimensional QCD on the sphere and on the Cylinder}, hep-th 9309107.}
\lref\REFgrwit{D. Gross and E. Witten, Phys. Rev. {\bf D21} (1980) 446.}
\lref\REFNeub{H. Neuberger, Nucl. Phys. {\bf B179} (1980) 253.}
\lref\REFwitten{E. Witten, Commun. Math. Phys. {\bf 141} (1991) 153\semi
J. Geom. Phys. {\bf 9} (1992) 303.}
\lref\REFgrmig{D. Gross and A. Migdal, Phys. Rev. Lett. {\bf 64}
(1990)127,  Nucl. Phys. {\bf B340} (1990) 333;
M. Douglas and S. Shenker,  Nucl. Phys. {\bf B335} (1990) 635
\semi E. Brezin and V. Kazakov, Phys. Lett. {\bf B236} (1990) 144.}
\lref\REFbessis{D. Bessis, C. Itzykson and J.-B. Zuber,
Adv. Appl. Math. {\bf 1} (1980) 109.}
\lref\REFalvarez{L. Alvarez-Gaum\'e, C. Gomez and J. Lacki, Phys. Lett.
{\bf B253} (1991) 56.}
\lref\REFbernard{G. 'tHooft, Phys. Rev. {\bf D14} (1976) 3432\semi
C. Bernard, Phys. Rev. {\bf D19} (1979) 3013.}
\lref\REFperiwal{V. Periwal and D. Shevitz, Phys. Rev. Lett. {\bf 64}
(1990) 1326; Nucl. Phys. {\bf B344} (1990) 731.}
\lref\REFNeuberger{H. Neuberger, Phys. Lett. {\bf B94} (1980) 199.}
\lref\REFhas{A. Hasenfratz and P. Hasenfratz,  Phys. Lett. {\bf B93} (1980)
165.}
\lref\REFdash{R. Dashen and D. Gross, Phys. Rev. {\bf D23} (1981) 2340.}

\Title{\vbox{\baselineskip12pt\hbox{PUPT-1459}
\hbox{hep-th/9404004}}}
{\vbox{\centerline{Instanton Induced Large $N$ Phase Transitions }\vskip0.15in
\centerline{in Two  and Four Dimensional QCD }\vskip0.15in
\centerline{}}}

\centerline{\vbox{\hsize3in\centerline{David J. Gross}
\smallskip\centerline{gross@puhep1.princeton.edu}}}
\bigskip
\centerline{and}
\bigskip
\centerline{
\vbox{\hsize3in\centerline{Andrei Matytsin}
\smallskip\centerline{matytsin@puhep1.princeton.edu}
}}
{\it
\bigskip\centerline{Department of Physics}
\centerline{Joseph Henry Laboratories}
\centerline{Princeton University}
\centerline{Princeton, NJ \ 08544}}

\vskip .5in

\centerline{\bf Abstract   }
\noindent
The $1/N$ expansion of the weak
coupling phase of two-dimensional QCD on a sphere is constructed.
It is  demonstrated  that the phase transition from the weak to the
strong coupling phase is induced by instantons.  A double scaling
limit of the theory at the point of the phase transition is constructed,
the value of string susceptibility is determined to be $\gamma_{str}=-1$.
The possibility of instanton induced large $N$ phase transitions in
four dimensional QCD is explored.
\Date{March 1994}

\secno 0
\newsec{Introduction}
A longstanding  problem has been  whether QCD has a
representation in terms of string theory.  Recently, this question
has been explored, and a string picture discovered, for QCD in two
dimensions \REFtaylor.
One uses the fact that the partition function and the Wilson
loops in two-dimensional QCD can be calculated exactly \REFmigdal,
\REFrusakov.
These known answers can then be interpreted in terms of sums over maps
between two-dimensional manifolds, yielding the string interpretation.
The hope is that some features of the two-dimensional theory, which
can be understood in detail due to its exact solvability, have their
counterparts in the real, four-dimensional world.

Apart from this, the two-dimensional Yang-Mills theory exhibits other
amusing features.  Douglas and Kazakov \REFdouglas\ considered QCD$_2$ on a
sphere and observed that it has a large $N$ phase
transition as one varies the area of the sphere. Unlike
other known large $N$ phase transitions \REFgrwit, which might be
regarded as lattice artifacts,  this one occurs  in the
continuum version of the theory. The physical reason for this
transition was not completely clear.

The string representation describes, by construction,  the strong coupling
phase of QCD. Remember
that in QCD$_2$ observables depend only on the product of the area with the
coupling so that strong coupling is equivalent to large area. It would be
interesting to understand what happens in the
weak coupling phase. Here one might conjecture  that the
string picture could break
down. Indeed there could be a phase transition and the weak coupling phase
might  correspond  to a
theory of point particles, rather than to a string theory. To understand this
phase
 it is necessary to construct the $1/N$ expansion for weak coupling.
This we shall do in the next section. We find that,
starting from the second term, all terms in the $1/N$ expansion vanish in this
phase. There are, however, nonperturbative, ${\cal O}(\ee^{-\gamma N})$,
corrections to the expansion. These corrections are due to the contribution of
instantons in  two-dimensional QCD. In section 3 we calculate the instanton
contribution
to the free energy. We find that while in the weak coupling phase
this contribution is exponentially small, it blows up as we approach the
phase transition point. The transition occurs when the entropy of instantons
starts dominating over their Boltzmann weight, $\ee^{-S_{\rm inst}}$.
The density
of instantons then goes from $0$, for weak coupling, to $\infty$, for strong
coupling\foot{Other large $N$ phase transitions, those that occur in matrix
models, or in the one-plaquette model of QCD can also be attributed to
instantons \REFNeub.}.

	Indeed, we can therefore argue that instantons are responsible for
confinement
in two dimensions. At first sight this might appear paradoxical. After all
confinement in QCD$_2$ is usually regarded as a perturbative effect. In fact in
any axial gauge the theory is quadratic and the gluon propagator gives rise to
a linear potential, which yields an area behavior for  any non-intersecting
Wilson loop   by summing Feynman diagrams. Thus one would say that the large
$N$ master field  is simply Gaussian fluctuations around the trivial, classical
vacuum $A_\mu =0$, and certainly does not describe instantons. However we note
that instanton contributions in QCD$_2$ are not easy to distinguish from
perturbative Gaussian fluctuations. To discuss instantons at all we must
compactify space-time, {\it i.e.} add the point at infinity. Consider QCD$_2$
on a sphere of area $A$. Then there are classical solutions of the Yang-Mills
equations which correspond to magnetic monopoles sitting at the center of the
sphere, giving rise to constant fields, $F_{\mu \nu}$ on the sphere. These have
quantized charges, $
F_{\mu \nu} \sim n/\tilde g A,$ are topologically distinct form the
perturbative vacuum and have action $S\sim n^2/ \tilde g^2A$. Therefore in the
limit of infinite $A$ these contributions are indistinguishable from  small
perturbations about zero field.
To ascertain whether they do contribute we should study the theory for finite
area, calculate the instanton contribution  and then take the infinite area
limit. As we shall see for small area the instantons do not contribute, but for
large areas they do, in fact they drive the Douglas-Kazakov transition.

Analogous large $N$ instanton
induced phase transitions might  occur  in QCD in
higher dimensions. We discuss the possibility of large $N$
phase transitions   for
four-dimensional QCD
and investigate whether they could invalidate a string representation for
short-distance physics. We present evidence that such transitions are unlikely.

Finally, in section 4 we explore the behavior of QCD$_2$ near the phase
transition,
show that there exist a doubling scaling limit near this point, construct this
limit explicitly and find the  corresponding ``string equation" (that is, the
equation which determines
the specific heat of the model).

\newsec{The Weak Coupling Phase}
As we have already mentioned,
the partition function of a
two-dimensional gauge theory can be calculated exactly for any manifold
${\cal M}$ of genus $G$ and area $A$.
The result can be expressed as a sum over
all irreducible representations of the gauge group \REFrusakov :
\eqn\pfgeneral{ \eqalign{
{\cal Z_{M}}&= \int [{\cal D\thinspace A}^{\mu}]{\rm e}^
{-{1\over {4 \tilde{g}^2}}\int d^2 x \sqrt{g}
\thinspace {\rm tr} F^{\mu \nu}F_{\mu \nu}} \cr
&= Z(G, \lambda A, N)=\sum_{R}({\rm dim} R)^{2-2G} {\rm e}^{-
{ \lambda A\over2 N}C_{2}(R)} \cr}
}
with ${\rm dim} R$ and $C_{2}(R)$ being the dimension and the quadratic
Casimir of the representation $R$ ($\lambda$ is related to $\tilde{g}$ by
$\lambda= \tilde{g}^2 N$).
Since the coupling constant and the area of the manifold enter ${\cal Z_{M}}$
only as a single combination $\lambda A$, we will put $\lambda=1$ in
what follows.
We will furthermore restrict ourselves to the case of the gauge group $U(N)$.

In this section we will analyze the $1/N$ expansion of the
partition function of two-dimensional QCD on the sphere,
\eqn\pfsphere{Z(G=0, A, N)=\sum_{R} ({\rm dim}R)^{2}
{\rm e}^{-{A\over 2 N}C_{2}(R)}.
}
If $A$ is kept fixed,
the free energy of the theory can be expanded as
a power series in $1/N$:
\eqn\fegeneral {F(G, A, N)\equiv {\rm ln}Z(G, A, N)
=\sum_{g=G}^{\infty} N^{2-2g}f_{g}^{G}(A) .}
We will show that for $A<A_{c}=\pi^2$
\eqn\f { f_{G=0}^{g}(A)= 0,}
for all $g \ge 2$. This means that the $1/N$ expansion in the phase of
small areas has only two terms. Furthermore, we will evaluate the exponential
correction to this trivial
$1/N$ expansion.

The partition function \pfsphere \ can be written explicitly as
\eqn\spfl{Z_{0}(A, N)=
\sum_{\scriptstyle l_1>\ldots >l_N} \Delta^2 (l_1, \ldots, l_N)\thinspace
{\rm e}^{-{A\over 2 N}C_{2}(l_1, \ldots, l_N)}.
}
We have taken into account that the representations $R$ of $U(N)$ can be
labeled by a set of integers $+\infty>l_1>l_2>\ldots >l_N>-\infty$,
the dimension of such representation being given by the Van der Monde
determinant
\eqn\vander{{\rm dim}R=\Delta(l_1, \ldots , l_N)\equiv
\left|\matrix{1 & \ldots & 1 \cr
	      l_1&\ldots & \l_N \cr
	      \vdots &\ldots&\vdots \cr
	      l_1^{N-1}& \ldots & l_N^{N-1} \cr }\right|
=\prod_{i<j}(l_i - l_j).
}
In terms of $\{l_j\}$
the Casimir operator $C_2(l_1, \ldots , l_N)$
has the following form:
\eqn\casimir{
C_2(l_1, \ldots, l_N)=\sum_{i=1}^{N}n_i(n_i-2i+N+1)=
{N\over 12}(N^2-1)+\sum_{i=1}^{N}\Bigl(l_i-{N-1\over 2}\Bigr)^2 ,
}
where $l_i=n_i-i+N$.
Note that \spfl \ is symmetric with respect to any permutation of
$l_j$ and vanishes when any two of $l_j, l_k$ are equal. Therefore one can drop
the condition $l_1>\ldots>l_N$, and, introducing $1/N!$, extend the summation
to all unrestricted sets of $\{l_1,\ldots ,l_N\}$:
\eqn\spfx{\eqalign{
Z_0 (A,N)=&{1\over N!} \ee^{-{A\over 24}(N^2-1)}\sum_{l_1,\ldots ,l_N =-\infty}
^{+\infty} \Delta^2 (l_1,\ldots, l_N)\thinspace\ee^{-{A\over 2N}\sum_{i=1}^{N}
\bigl(l_i-{N-1
\over 2}\bigr)^2}\cr
=&{1\over N!} \ee^{-{A\over 24}(N^2-1)}\sum_{x_1,\ldots ,x_N=-\infty}
^{+\infty} \Delta^2 (x_1,\ldots, x_N)\thinspace
\ee^{-{A\over 2N}\sum_{i=1}^{N}x_i^2}\cr}
}
where we have introduced
new summation variables $x_i=l_i-{N-1\over 2}$.

In this form  $Z_0 (A,N)$ can be evaluated with the aid of
orthogonal polynomials \REFbessis, which have been widely applied in
the random matrix models. The new effects come from the discreteness of $x_i$,
which can assume only integer values\foot{If $N$ is even,
$x_i$ will be, strictly speaking, half--integers. This does not alter our
conclusions.}.

We introduce a set of polynomials in $x$,
$p_n(x|\alpha)=x^n+\ldots$\
(where we have explicitly indicated the dependence on the parameter $\alpha
\equiv A/2N$)
orthogonal with respect to the following discrete measure:
\eqn\orth{\langle p_n|p_m\rangle \equiv \sum_{x\in \IZ}\ee^{-\alpha x^2}
p_n(x|\alpha) p_m(x|\alpha) =\delta_n^m h_n(\alpha).
}
Adding to any row of the Van der Monde determinant \vander \ any linear
combination of the previous rows  does not change the determinant.
This allows us to
represent $\Delta(x)$ as
\eqn\vdm{\Delta(x_1 \ldots x_N)= \det \bigl|x_i^{j-1}\bigr|_{i,j=1}^N
=\det\bigl|p_{j-1}(x_i|\alpha)\bigr|_{i,j=1}^N
}
Then, we
expand the determinants in \spfx , and use
\orth \ to carry out the summation over $x_i$.  The result is
\eqn\pfh{
Z_0 (A,N)={1\over N!}\ee^{-{A\over 24}(N^2-1)}\prod_{j=0}^{N-1}h_n
\Bigl(\alpha=
{A\over 2N}\Bigl).
}
To calculate $h_n (\alpha)$ we shall now derive a set of recursion relations
they must satisfy. The key idea is that, by construction, the polynomial
$p_n$ is orthogonal to any polynomial of degree less than $n$.
This, in turn, implies the relation
\eqn\xop{x p_n(x|\alpha)=p_{n+1}(x|\alpha)+ S_n(\alpha)p_n(x|\alpha)+
R_n(\alpha)p_{n-1}(x|\alpha).
}
Taking the scalar product of both sides of \xop \ with $p_{n-1}$ one easily
derives
$R_n(\alpha)=h_n (\alpha)/h_{n-1}(\alpha)$. The same way, it is not
difficult to see that $p_{n-2}$ and lower degree polynomials
do not enter the right hand side of \xop. Also, due to the symmetry
of our measure with respect to $x\rightarrow -x$, $S_n(\alpha)\equiv 0$.

Therefore, it would suffice to derive
the relations between $R_{n+1}(\alpha)$ and $R_n(\alpha)$.
To this effect we differentiate the equation
\eqn\norm{
h_n (\alpha)=\langle p_n | p_n \rangle \equiv
\sum_x \ee^{-\alpha x^2}p_n^2(x|\alpha)
}
to get
\eqn\recrel{\eqalign{
{d h_n(\alpha)\over d\alpha}=&\sum_x \ee^{-\alpha x^2}\biggl[2 p_n(x|\alpha)
{\partial p_n(x|\alpha) \over \partial \alpha} - x^2 p_n^2(x|\alpha)\biggr] \cr
=&-\sum_x \ee^{-\alpha x^2}x^2 p_n^2(x|\alpha)=-\langle x p_n |x p_n \rangle
\cr
=&-\langle p_{n+1}+R_n p_{n-1}|p_{n+1}+R_n p_{n-1} \rangle \cr
=&-(h_{n+1}+R_n^2 h_{n-1}) = - h_n ( R_{n+1} + R_{n}) \cr}.
}
We have used the fact that $\langle p_n|{\partial p_n\over \partial \alpha}
\rangle =0$ because ${\partial p_n\over \partial \alpha}$ is a polynomial
of the order $n-2$ and is orthogonal therefore to $p_n$.
Finally, we obtain our recursion relations
\eqn\rel{\Biggl\{\eqalign{&
R_{n+1}(\alpha)= -\Bigl(R_n(\alpha)+ {d \over d\alpha}\ln h_n(\alpha)\Bigr)
\cr
&h_{n+1}(\alpha) = R_{n+1}(\alpha) h_n (\alpha)\cr
}
}
They have to be supplemented by the obvious initial conditions
\eqn\init{
h_0(\alpha)=\sum_x \ee^{-\alpha x^2} ,\quad  R_0(\alpha)=0 .
}
One can eliminate $h_n(\alpha)$ to get a closed form of the  relation
for $R_n(\alpha)$:
\eqn\volterra{ R_{n+1}(\alpha) - R_{n-1}(\alpha)=- {d \over d \alpha}
\ln R_n(\alpha).
}
When viewed as an evolution equation for functions $\ln R_n(\alpha)$, $\alpha$
being treated as time, \volterra \ is called the Volterra equation \REFalvarez.

Once $R_n(\alpha)$ are known,
\eqn\hn{
h_n(\alpha)=h_0(\alpha) \prod_{i=1}^n R_i(\alpha).
}

Now we would like to find $R_n(\alpha= A/2N)$ for all $n$. Let us notice that
$\alpha\rightarrow 0$ in the $N \rightarrow \infty$ limit.
This suggests that the leading term of the $1/N$ expansion can be obtained
if we approximate the sum in \init \ by an integral.
Remarkably, such approximation gives us not merely the first term but
the whole $1/N$ expansion, all corrections being exponentially small.
Indeed, we can elucidate this using the Poisson resummation formula: if
$$ F(y) ={1\over 2\pi}\int_{-\infty}^{+\infty}f(x)\ee^{-i y x }dx \, , $$
then
\eqn\poisson{
\sum_{n=-\infty}^{+\infty}F(n)=\sum_{m=-\infty}^{+\infty}f(2\pi m).
}
Choosing
$$ f(x)=\ee^{-{A\over 2N}x^2} \ ,$$
we derive
\eqn\hresummed{
h_0\Bigl(\alpha={A\over 2N}\Bigr)=\sqrt{2\pi N \over A} \sum_{n\in \IZ}
\ee^{-{2\pi^2 n^2
\over A}N}
=\sqrt{2\pi N \over A} \big(1+ 2 \ee^{-{2\pi^2\over A}N} +\ldots \big).
}
Let us ignore the exponential correction for a moment, then $h_0^{(0)}({A/ 2N})
=\sqrt{2\pi N / A}$, and the recursion relations are easy to solve with the
result
\eqn\hermite{
R_n^{(0)}\Bigl({A\over 2N}\Bigr)= {N n \over A} \ , \qquad
h_n^{(0)}\Bigl({A\over 2N}\Bigr)=\sqrt{2\pi}
n! \Bigl({N\over A}\Bigr)^{n-\half}\, .
}
In fact,
these $R_n$ correspond to the set of Hermite polynomials $H_n(x)$, orthogonal
with respect to the continuous measure, as if there were no discrete
structure at all:
\eqn\hermeasure{
\int_{-\infty}^{+\infty} \ee^{-\alpha x^2} H_n(x|\alpha)H_m(x|\alpha) dx
=\delta_n^m { \sqrt{2 \pi}n! \over(2\alpha)^{n-\half}} \, .
}
The discreteness of our true measure \orth \ reveals
itself in the exponentially small
corrections to $h_0(\alpha)$. This suggests that $h_n(\alpha)$ may be given
by \hermite \  without any $1/N$ corrections. We will see that this is indeed
the case provided that the number of the polynomial $n$
does not exceed a certain critical value, namely,
\eqn\trans{
n<n_{cr}=N{\pi^2 \over A}.
}
Technically speaking, the reason is that
in the process of iteration of \rel \ the
exponential error accumulates and can grow
sufficiently large. This leads to a  phase
transition in the structure of the discrete orthogonal polynomials
at $n=n_{cr}$.
In fact, when $n>n_{cr}$, $p_n(x|\alpha)$ differ drastically from  Hermite
polynomials and \hermite \ is no longer a good approximation.
To see how this transition occurs, let us trace the contribution from
the first correction to $h_0(\alpha)$. If we assume the ansatz
\eqn\ansatz{
R_n\Bigl({A\over 2N}\Bigr)= {N n \over A}+ c_n (A) \ee^{-{2\pi^2 N\over A}}
}
where $$c_n (A) \ee^{-{2\pi^2 N\over A}} \ll {N n \over A}$$
we will have
$$\ln R_n\Bigl({A\over 2N}\Bigr) = \ln \Bigl({N n \over A}\Bigr)
+ {A \over Nn}
c_n (A) \ee^{-{2\pi^2 N\over A}} +\ldots .
$$
As a consequence of \volterra \  one can
derive the recursion relation for $c_n (A)$,
\eqn\relcnA{
c_{n+1}(A)-c_{n-1}(A)=-{2\over n}\biggl[{2\pi^2 N \over A} c_n (A) +
{d\over dA}\big(A c_n (A)\big)\biggr].
}
The initial conditions for these coefficients are
$$ c_0(A)=0 \ , \qquad c_1(A)=-{8\pi^2 N^2 \over A^2}. $$
To solve for $c_n(A)$,
it will be convenient to introduce a parameter $\xi ={2\pi^2 N \over A}$
and to define the new unknown functions $G_n(\xi)$ such that
\eqn\cg{
c_n(A)=-{2\over \pi^2}\xi^2 G_n(\xi).
}
Then
\eqn\grel{
G_{n+1}(\xi)- G_{n-1}(\xi)= -{2\over n}\Bigl[(\xi-1)G_n(\xi) -
\xi G_n^{\prime}(\xi)\Bigr]
}
and one can easily see that $G_n(\xi)$ is a polynomial of the order $n-1$:
$$ G_1(\xi)=1 \ ,\  G_2(\xi)=-2(\xi-1) \ ,\  G_3(\xi)=2 \xi^2 - 6 \xi + 3 \
 , \;  etc. $$
We can find $G_n(\xi)$ for an arbitrary $n$
by considering the generating functional
$$F(\xi,z)= \sum_{n=1}^{\infty}z^n G_n(\xi)$$
which, by virtue of \grel , satisfies the partial differential equation
\eqn\pde{
(1-z^2){\partial F\over \partial z} -2 \xi{\partial F\over \partial\xi}
=\biggl[z+{1\over z} -2 (\xi -1)\biggr]F.
}
The function $F(\xi, z)$ is fixed uniquely by the requirement that as
$z\rightarrow 0$ ,$$F(\xi, z) = z+ {\cal O}(z^2).$$
It is not difficult to demonstrate that it is given by
\eqn\F{
F(\xi, z)={z\over (1-z)^2}\ee^{-{2\xi z\over 1-z}}.
}
Then all $G_n(\xi)$ can be determined in terms of contour integrals
$$G_n(\xi)=\oint_{C_z} {dz \over 2\pi i}{1\over z^{n+1}}F(\xi, z)\, ,
$$
where the contour $C_z$ encircles the point $z=0$ in such a way that
$z=1$ is outside of the
contour.

Let us cast this integral into a more convenient form.
Introducing the new integration variable
$$t={z\over 1-z}$$
we can represent it as
\eqn\contint{
G_n(\xi)=\oint_{C_t}{dt \over 2\pi i} \ee^{-2\xi t}\Bigl(1+{1\over t}\Bigr)^n.
}
The contour $C_t$ must encircle $t=0$ and pass to the right of $t=-1$.
Now, expanding $(1+{1\over t})^n$ in powers of $1/t$ we get an explicit
expression
\eqn\polynomial{
G_n(\xi)= \sum_{k=0}^{n-1}{(-2)^k \over k!}{n!\over (k+1)! (n-k-1)!}\xi^k.
}
In fact,
$$G_n(\xi)= L_{n-1}^{(1)}(2\xi)$$
where $ L_{n-1}^{(1)}(x)$ are the so-called generalized Laguerre polynomials,
defined as the coefficients of the expansion
\eqn\laguerre{
\sum_{n=0}^{\infty} L_{n}^{(p)}(x) z^n={1\over (1-z)^{p+1}}\ee^{
-{x z \over 1-z}}.
}
Combining the pieces we find the answer for $R_n$:
\eqn\Rncorr{
R_n\Bigl({A\over 2N}\Bigr)= {N n \over A} - {2\over \pi^2}\Bigl(
{2\pi^2 N\over A}\Bigr)^2
L_{n-1}^{(1)}\Bigl({ 4\pi^2 N\over A}\Bigr) \ee^{-{2\pi^2 N\over A}}.
}
Let us emphasize that the second term in \Rncorr \ must be much less that
the first one, otherwise our calculation becomes invalid. Evidently,
the first term does dominate for any finite fixed $n$ as
$N \rightarrow \infty$. However, {\it a priori}, if $n\sim {\cal O}(N)$,
\Rncorr \ may or may not be correct. Indeed,
since $L_{n-1}^{(1)}(x)$ is a polynomial of order $n-1$, it contains terms
of order $(4 \pi^2 N /A)^{n-1}$ which at $n\sim N$ can win over the small
factor $\ee^{-2\pi^2 N / A}$ thus changing the behavior of $R_n({A/2N})$.
Still, we need to know $R_n(A/2N)$ for $n$ up to $n=N-1$ to be able
to calculate the partition function of the two-dimensional QCD \pfh .
Therefore, we need to
investigate the asymptotics of \Rncorr \ as $n\sim N
\rightarrow \infty$ in more detail.

To this effect note that the integral in \contint \ can be evaluated by
the steepest descent method. Indeed, if $n$ is of order $N$,
$$ G_n(\xi)= \oint_{C_t}{dt \over 2\pi i} \ee^{-N \Phi(t)} $$
where $\Phi(t)$ is a functional which assumes values of order 1,
$$\Phi(t)={4\pi^2\over A}t- {n\over N}\ln \Bigl(1+{1\over t}\Bigr) .$$
It is easy to see that the saddle points of $\Phi(t)$ are solutions
of the quadratic equation
$$t^2+t+{A\over 4\pi^2}{n\over N}=0 .$$
Recalling the notation, introduced in \trans \ we find
\eqn\tplusminus{
t=t_{\pm}=-\half \pm \half\sqrt{1-{n\over n_{cr}}}.
}
We see that the position of the saddle points in the complex plane is
qualitatively
different for $n<n_{cr}$ and $n>n_{cr}$. In the first case it is possible to
deform
the contour as shown in the figure\foot{Please note that
the contours in Fig.1 are directed {\it downwards}, because the contour $C_t$
is assumed to go counterclockwise around $t=0$.} so that
the integral is determined
by the contribution of the rightmost saddle point, $t=t_{+}$:
\ifig\saddle{Lines of equal level for ${\rm Re}\ \Phi(t)$ and the
integration contour in the two
cases: \quad (left)\  $n<n_{cr}$ \qquad\  and \quad (right)\  $n>n_{cr}$.}
{\epsfxsize 2.0in\epsfbox{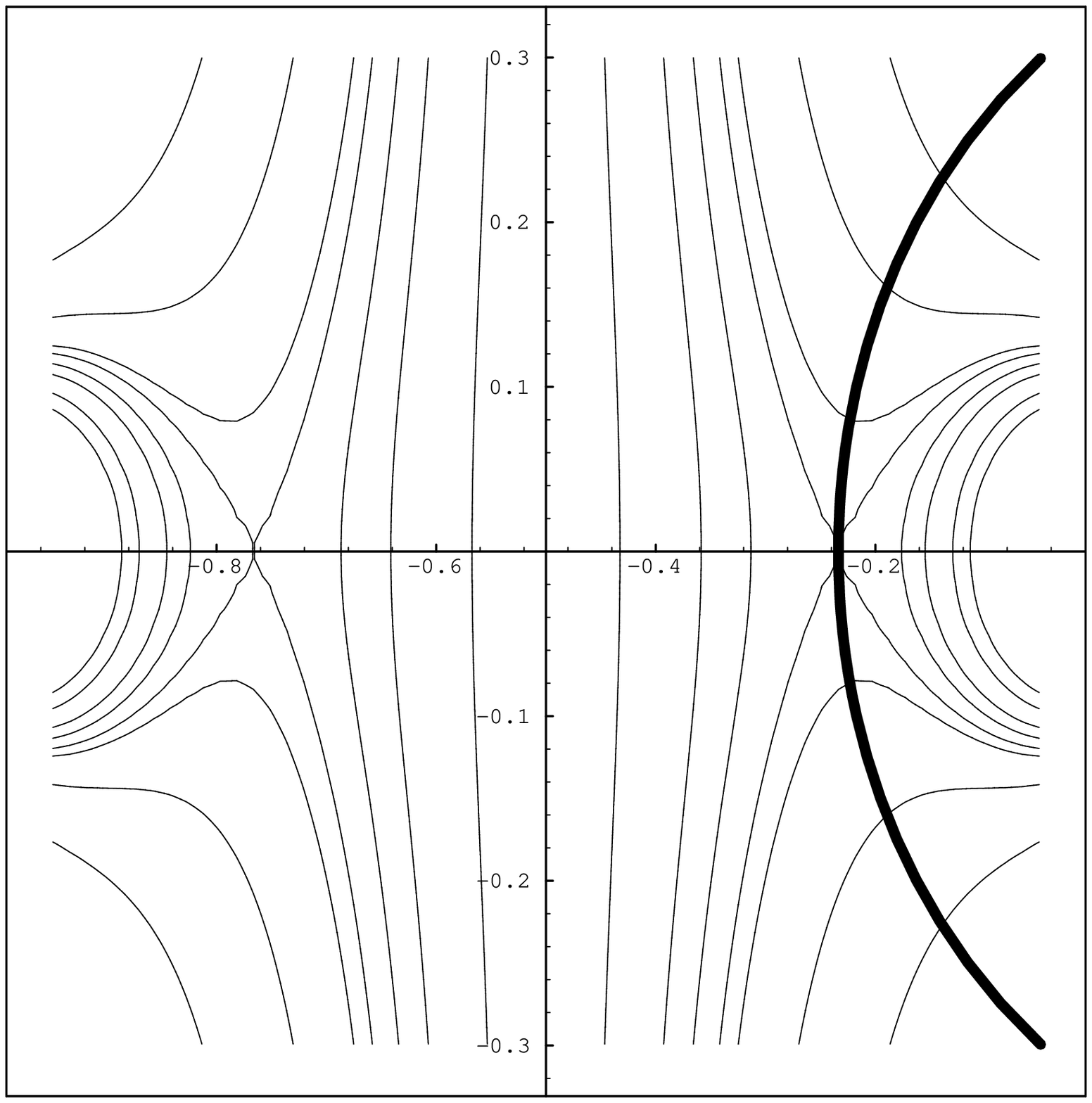}\hskip0.5in
 \epsfxsize 2.0in\epsfbox{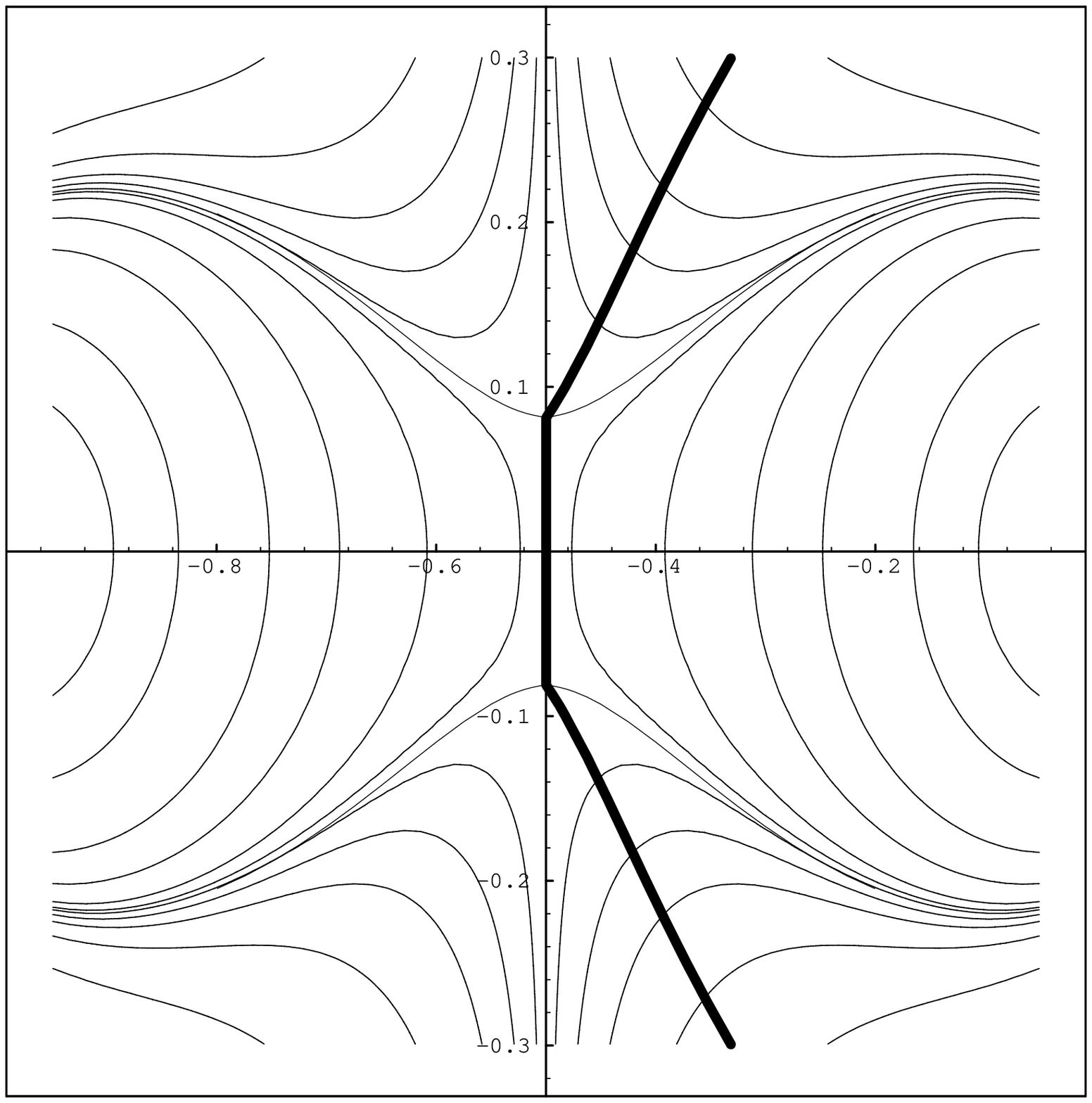}}
\eqn\gnasympt{
G_n(\xi)\simeq \ee^{-N\Phi(t_+)}\int_{\tau=-i\infty}^{\tau=+i\infty}
{d\tau \over 2\pi i}\ee^{\half N \Phi^{\prime \prime}(t_+) (i\tau)^2}
=-{1\over \sqrt{2\pi}}{\ee^{-N\Phi(t_+)}
\over\sqrt{N|\Phi^{\prime \prime}(t_+)|}}.
}
Now, after some algebra one finds
$$\Phi^{\prime \prime}(t_+)=-{n\over N}{2t_+ +1\over t_+^2 (t_+ +1)^2}=
-{16 n_{cr}^2 \over nN}\sqrt{1-{n\over n_{cr}}}<0
$$
and
$$\Phi(t_+)=-{2\pi^2 \over A} +{2\pi^2 \over A}\Biggl[\sqrt{1-{n\over n_{cr}}}
-{1\over 2}{n\over n_{cr}}\ln{1+ \sqrt{1-{n\over n_{cr}}}\over
1-\sqrt{1-{n\over n_{cr}}}}\Biggr].
$$
The quantity we are interested in is, in fact, the combination entering
our formula for $R_n$,
\eqn\gresult{
G_n(\xi)\ee^{-{2\pi^2 N \over A}}\simeq (-)^{n-1} {1\over 4}{1\over \sqrt{2\pi
n_{cr}}}
\sqrt{n\over n_{cr}}\Bigl(1-{n\over n_{cr}}\Bigr)^{-1/4}
\ee^{-{2\pi^2 N \over A}
\gamma({n\over n_{cr}})}
}
where\foot{The $(-)^n$ comes from the fact that $1+{1\over t_+}<0$ and thus
$\Phi(t_+)$ has, strictly speaking, an imaginary part, equal to $i\pi n/N$.}
\eqn\gammafunction{
\gamma(x)=\sqrt{1-x}- {x\over 2}\ln {1+\sqrt{1-x}\over 1-\sqrt{1-x}}
=2\sqrt{1-x} \sum_{s=1}^{\infty}{(1-x)^{2s}\over 4s^2-1} >0 .
}
If $x=n/n_{cr}<1$ then always $\gamma(x)>0$. This implies that \gresult, and
hence the correction in \Rncorr, is indeed exponentially small, ${\cal O}
(\ee^{-{2\pi^2 N \over A} \gamma})$.

Substituting \gresult \ into \Rncorr \ we get
\eqn\rass{
R_n= {N n \over A} + (-)^n
\sqrt{2 n\over \pi^5}n_{cr}\Bigl(1-{n\over n_{cr}}\Bigr)^{-1/4}
\ee^{-{2\pi^2 N \over A}\gamma({n\over n_{cr}})} + \ldots.
}
This asymptotic formula is applicable for $0<n<n_{cr}$.

Now it is easy to see how the phase transition in two-dimensional QCD
occurs. By virtue of \pfh \ the partition function depends only on $R_n$
with $0<n<N$. So, if $N<n_{cr}$, all relevant $R_n$ are given
by \rass \ and the system is in its weak coupling phase. The condition on the
critical point is $n_{cr}=N$, which amounts to $A_{cr}=\pi^2$.
This is precisely the value found by Douglas and Kazakov.

Finally, let us evaluate the free energy of QCD in the weak coupling phase.
Using \pfh, \hn \ and \ansatz \ we easily get
\eqn\free{\eqalign{
F(A,N)\equiv&\ln Z_0 (A,N)\cr=&-{A\over 24}(N^2-1)+N\ln h_0 \Bigl({A\over 2N}
\Bigr) +
\sum_{i=1}^{N-1} (N-i)\ln R_i \Bigl({A\over 2N}\Bigr)\cr
=&-{A\over 24}(N^2-1) - {N^2\over 2}\ln A +
\biggl[2N +{A\over N}\sum_{i=1}^{N-1} {N-i\over i}c_{i}(A)\biggr]\ee^{-\xi}.\cr
}}
Using \cg \ and the contour integral representation \contint \ one
can, after some manipulations with the contour integral,  reduce it to
\eqn\freefinal{
F(A,N)=-{A\over 24}(N^2-1) - {N^2\over 2}\ln A + 2\ee^{-\xi}G_N (\xi).
}
Using \gresult \ we obtain our result
\eqn\instantonenergy{
F(A,N)=-{N^2\over 2}\Bigl({A\over 12} +\ln A\Bigr) + {A\over 24}
-{(-)^N\over \sqrt{2\pi N}}{A\over 2\pi^2}\Bigl(1-{A\over \pi^2}\Bigr)^{-1/4}
\ee^{-{2\pi^2 N\over A}\gamma({A\over \pi^2})} +\ldots .
}

A brief inspection of this formula shows that the $1/N$ expansion is extremely
simple: all its terms, except for the first two, vanish\foot{The
${\cal O}(N^0)$ term, $A/24$, is simply the factor in front
of the sum \spfx. It does not have any physical significance.}. All non-trivial
effects are incorporated in the nonperturbative corrections to the free energy,
which are invisible within the $1/N$ expansion. Still, it would make
sense to have a clear physical picture of these corrections, as well as
the phase transition. A natural hypothesis is that the ${\cal O}(\ee^{-{
2 \pi^2 N\over A}\gamma})$ term in \instantonenergy \ is due to the
contribution of instantons to the free energy, while the $1/N$ expansion
(finite in this case) represents the energy of quantum fluctuations around
the vacuum of perturbation theory.  We shall investigate this
possibility below.

\newsec{The Contribution of Instantons}

In this section we shall evaluate   the
contribution of instantons to the partition function. We recall Witten's
analysis of  QCD$_2$ \REFwitten,
where he    generalized the Duistermaat-Heckman theorem
to the non-Abelian case and showed that the
partition function is given by a sum over contributions localized at classical
solutions of the theory. For finite $N$  Witten showed that the $SU(N)$ path
integral is given by a sum over  unstable instantons, where each instanton
contribution is given by a finite, but non-trivial, perturbative expansion. The
order of perturbation theory required grows with $N$, and as $N \to \infty$ one
gets contributions from all orders, even though the theory is so trivial. We
will be able to  indirectly evaluate and sum this infinite perturbative
expansion about the instantons.

First, let us classify the instantons of the two-dimensional gauge theory on a
sphere. We are still working with the gauge group $U(N)$. By
instantons we mean    solutions of the classical field equations
$$\partial^{\mu}F_{\mu \nu} + i\thinspace [A^{\mu}, F_{\mu \nu}]=0 $$
$$F_{\mu \nu}= \partial_{\mu}A_{\nu}-\partial_{\nu} A_{\mu}+
i \thinspace [A_{\mu}, A_{\nu}] $$
which are not gauge transformations of the trivial solution, $A_{\mu}\equiv 0$.
All such solutions (up to a gauge transformation) are simply
\eqn\instanton{
A_{\mu}(x)=\left(\matrix{n_1 {\cal A}_{\mu}(x) & 0 & \ldots & 0 \cr
			 0      & n_2 {\cal A}_{\mu}(x) & \ldots & 0\cr
			 0 &\ldots&\ldots&0\cr
			 \vdots&\vdots&\ddots&\vdots\cr
			 0& 0 &\ldots &n_N{\cal A}_{\mu}(x)\cr
}\right)
}
where ${\cal A}_{\mu}(x)={\cal A}_{\mu}(\Theta, \phi)$ is the Dirac
monopole potential,
$${\cal A}_{\Theta}(\Theta, \phi)=0 , \quad\   {\cal A}_{\phi}
(\Theta, \phi)={1-\cos \Theta\over 2}, $$
$\Theta$ and $\phi $ being the polar (spherical) coordinates on $S^{2}$;
and $n_1, \ldots, n_N$ are arbitrary integers\foot{The condition that $n_i$
can assume only integer values follows from the Dirac quantization condition
for magnetic monopole,
which is easy to generalize to our case.}.
The action corresponding to such instanton is straightforward
to calculate\foot{We keep in mind
our convention that $\lambda\equiv {\tilde g}^2 N =1 $.}
\eqn\Sinst{
S_{\rm inst}(n_1, \ldots , n_N)={1\over 4 {\tilde g}^2}\int d^2 x \thinspace
{\rm tr}
F^{\mu \nu} F_{\mu \nu} ={2\pi^2\over {\tilde g}^2 A}\sum_{i=1}^{N}n_i^2
={2\pi^2 N\over A}\sum_{i=1}^{N}n_i^2.
}
These instantons are unstable. For $SU(N)$ there are $2N-2$ negative modes.
These will produce a factor of $i^{2N-2}=(-1)^{N-1}$ in the determinant of
fluctuations about  the instanton. Nonetheless the {\it exact} partition
function of two-dimensional QCD is given by the sum over all instanton
contributions $w(n_1, \ldots , n_N)$:
\eqn\witten{
{\cal Z}=\sum_{n_1,\ldots ,n_N=-\infty}^{+\infty}w(n_1, \ldots, n_N)
\thinspace\ee^{-S(n_1,\ldots ,n_N)}
}
where $w(n_1, \ldots, n_N)$ are determined  a perturbative expansion about
  the instanton. It would be hopeless to evaluate these weights directly, but
they can be found from the exact answer \spfx , by rewriting the sums over the
lengths of the rows of the Young tableaux using the Poisson resummation formula
(note that the magnetic charges are dual to the lengths of the rows).
For the gauge group $U(N)$ this resummation has been performed   explicitly by
Minahan and Polychronakos \REFminahan\foot{A similar resummation formula
was  used by D'Adda {\it et al} \REFDADDA\ to argue, as we shall below, that
the  large $N$ phase transition is due
to topologically non-trivial configurations.},  who find
\eqn\minahan{
w(n_1, \ldots, n_N)=\int_{-\infty}^{+\infty}\prod_{i=1}^{N}
\thinspace\prod_{i<j=1}^{N}
\big(y_{ij}^2- 4\pi^2 n_{ij}^2\big)\thinspace
\ee^{-{N\over 2A}\sum_{i=1}^{N}y_i^2}
}
where
$$y_{ij}\equiv y_i-y_j \, , \quad n_{ij}\equiv n_i-n_j \, . $$

Let us calculate the contribution of configuration involving only
one single instanton of  charge $n=1$, $w_1\equiv w({\scriptstyle
n_1=1;\thinspace n_2=\ldots=n_N=0})$,
and compare in to the contribution of
the perturbative vacuum, $w_0 \equiv w({ \scriptstyle n_1=n_2=\ldots=n_N=0})$
in the large $N$ limit.
{}From \minahan \
\eqn\single{\eqalign{
w(1, 0, \ldots, 0)&=(-)^{N-1}\int_{-\infty}^{+\infty}dy_1
\ee^{-{N\over 2A}y_1^2}\cr
&\int_{-\infty}^{+\infty}\prod_{i=2}^{N}dy_i \thinspace
\Delta^2 (y_2, \ldots, y_N)\thinspace
\ee^{-{N\over 2A}\sum_{i=2}^{N}y_i^2}
\prod_{i=2}^N \big[4\pi^2- (y_i -y_1)^2\big].\cr}
}
We can do the integral over $y_2, \ldots, y_N$ by the saddle point method. In
fact as far as these variables are concerned we have a ordinary Gaussian matrix
model of a $N-1$ dimensional matrix, whose eigenvalues are $y_2, y_3, \dots,
y_N$.
The saddle point values of these variables are of order 1
and have their distribution determined by the ${\cal O}(N^2)$ term in the
exponential, ${N\over 2A}\sum y_i^2$, thus the density of eigenvalues,
$\rho(y)$ is given by$$\rho(y)={1\over \pi}\sqrt{{1\over A} -{y^2\over 4A^2}}
.$$
The product
\eqn\product{
\prod_{i=2}^N \big[4\pi^2- (y_i -y_1)^2\big]={\rm exp}\Bigl\{
\sum_{i=2}^{N}{\rm ln}\big[4\pi^2- (y_i -y_1)^2\big]\Bigr\}
}
would modify the exponential by a ${\cal O}(N)$ term, which is
negligible compared to ${\cal O}(N^2)$ term, thus not changing $\rho(y)$.
Therefore, in the large $N$ limit we can just substitute the saddle point
values of $y_i$, $i=2, \ldots, N$ into \product,  to obtain
\eqn\average{
\eqalign{
&{1\over Z_{N-1}}\int_{-\infty}^{+\infty}\prod_{i=2}^{N}dy_i \thinspace
\Delta^2 (y_2, \ldots, y_N)\thinspace \ee^{-{N\over 2A}\sum_{i=2}^{N}y_i^2}
\prod_{i=2}^N \big[4\pi^2- (y_i -y_1)^2\big] \cr
&=\Bigl\langle \prod_{i=2}^N \big[4\pi^2- (y_i -y_1)^2\big] \Bigr\rangle
_{\rho(y)}   = {\rm exp}\Bigl[
{N\int dy \rho(y)\thinspace{\rm ln}( 4\pi^2- (y- y_1)^2)}\Bigr] , \cr
}
}
where\foot{To evaluate $Z_{N-1}$ we use Hermite orthogonal polynomials
\hermite, \hermeasure \ for $\alpha = N/2A$.}
\eqn\formula{
\eqalign{
Z_{N-1}=& \int_{-\infty}^{+\infty} \prod_{i=2}^{N}dy_i \thinspace
\Delta^2 (y_2, \ldots, y_N)\thinspace \ee^{-{N\over 2A}\sum_{i=2}^{N}y_i^2}
\cr
=&\prod_{n=0}^{N-2} h_0^{(0)} \Bigl(\alpha= {N\over 2A}\Bigr) =
\prod_{n=0}^{N-2}
\biggl[ \sqrt{2\pi} n! \Bigl({A\over N}\Bigr)^{n-\half}\biggr].
\cr}
}
To integrate in \single \ over $y_1$ we notice that  the contribution
to the integral over $y_1$
comes from the region $y_1 \sim 1/\sqrt{N}$, due to the presence of
$\ee^{- N y_1^2 /2A}$. So we can replace $ y_1   $ by $0$ in \average\
to get
$$ w_1=Z_{N-1} \biggl[\int_{-\infty}^{+\infty} \ee^{-{N\over 2A}
y_1^2}\thinspace d y_1 \biggr] \ee^{N\int dy \rho (y)
\thinspace {\rm \ln} (4\pi^2- y^2)}
$$
Also, comparing \minahan \ for $n_1=\ldots=n_N=0$ to \formula, we see that
$$w_0 = Z_{N}=
\prod_{n=0}^{N-1}
\biggl[ \sqrt{2\pi} n! \Bigl({A\over N}\Bigr)^{n-\half}\biggr]. $$
Then, the ratio of weights behaves at large $N$ as
\eqn\ratio{
{w_1 \over w_0} \simeq {(-)^{N-1}\over (N-1)!}\Bigl( {N\over A} \Bigr)^{N-2}
\ee^{N\int dy \rho (y) \thinspace {\rm \ln} (4\pi^2- y^2)}.
}
Evaluating the integral and using the Stirling's formula,
$${1\over (N-1)!}\Bigl( {N\over A} \Bigr)^{N-2}\simeq \ee^{N(1 -{\rm \ln} A)}$$
we obtain
$${w_1 \over w_0}\simeq (-)^{N-1} \ee^{{2\pi^2 \over A}N - {2\pi^2 \over A}N
\gamma({A\over \pi^2})},
$$
where $\gamma(x)$ is precisely the function introduced in \gammafunction.

Therefore the contribution of the one instanton configuration, relative to
the contribution of vacuum fluctuations, is given by
\eqn\relative{
{w_1 \over w_0}\ee^{-S_{\rm inst}(n=1)}
\simeq (-)^{N-1}\ee^{-{2\pi^2 \over A}N
\gamma({A\over \pi^2})}.
}
This result is in agreement\foot{To evaluate
the factor in front of the exponential
in \relative \ one needs a more subtle technique, which would be essentially
identical to the one we have developed in the previous section.}
with \instantonenergy.

We see that the perturbative vacuum dominates in the partition function
in the weak coupling phase. The transition at $A=\pi^2$  occurs when the
statistical weight of instantons grows large enough to make them
favorable configurations.  If we were to evaluate the expectation value of a
Wilson loop
of area $A_1$ on  a  sphere of area $A$ we would find two different results,
depending on whether $A>\pi^2$ or $A<\pi^2$.  One can say that the large $N$
master field in these two case is different. The small area phase would give a
Wilson loop that does not confine as we let   $A$ tend to infinity.  Thus the
perturbative master field does not give confinement. The correct, confining,
behavior of the Wilson
loop on the plane  is obtained by taking the large area phase and then letting
$A\to \infty$. We thus conclude that confinement is a consequence of the large
$N$ master field containing instanton contributions.

\newsec{Instanton Induced Phase Transitions.}

We have seen that in evaluating the free energy on a sphere of area $A$ for
QCD$_2$
the   instanton  density is zero ({\it i.e.}
 $\lambda^\infty$, with $ \lambda <1$) for   $A< \pi^2$,   and   is infinite
({\it i.e.}
 $\lambda^\infty$, with $ \lambda >1$) for $A>\pi^2$.
The jump in the instanton density from zero to infinity
induces a phase transition.
The possibility of such
an instanton induced transition is present in any large $N$ theory.
The instanton
density is always suppressed exponentially as $N\to \infty$ due to the fact
that the instanton action
behaves as $S_{\rm inst} \sim c_1/g^2 \sim c_1N/\lambda$, with $\lambda=g^2
N$. However, the fluctuations about the instanton can give rise to an
{\it entropy\/}\  factor that can overcome the
Boltzman weight $\exp[- S_{\rm inst}]$. The dominant  term is the contribution
of zero modes, which produce  a factor of  $[c_2/\lambda]^{N_z\over 2}$,
where $N_z$ is the number of zero modes, one inverse
coupling for each zero mode. These appear  from the normalization of the
collective coordinates in the collective coordinate Jacobian. If $N_z\sim
2 c_3N$, which is the case in QCD
(both in two and in four dimensions), then  the
density of instantons is proportional to
$ [(c_2 / \lambda)^{c_3} \exp(-c_1/ \lambda)]^N$.
This density will blow up, as long as $c_2 c_3/c_1>e$,  for $\lambda$
greater than the
smallest solution of $ c_1/\lambda = c_3 \log[ c_2/\lambda] $.

In our case of ${\rm QCD}_2$ the number of zero
modes of the instanton, {\it i.e.}
the number of generators of $U(N)$ that do not
commute with the instanton field,
is $N^2 -(N-1)^2 -1= 2N-2$. Thus $c_3=1$.
Also $c_1=2 \pi^2$ and $c_2= 4 \pi^2 e$. The above criterion
is satisfied, namely
$ c_2 c_3/c_1 =2e >e$, and therefore a phase transition will take place.
Therefore, if this was the exact instanton density,
the transition would occur at $A/\pi^2 =0.746729$.
More generally the instanton density is given by
$[(c_2 / \lambda)^{c_3} \exp(-c_1/ \lambda +f(\lambda))]^N $,
where $f(\lambda)$ arises from the perturbative expansion  about the instanton.
In our case we were able to evaluate this function exactly,
see \instantonenergy,
and determine the exact instanton density to be
\eqn\density{ \rho(A) =  (-)^{N-1} \sqrt{2 \over \pi N}
\left(1-{A\over \pi^2}\right)^{-{1\over 4}}
 \left({4\pi^2 \over A}\right)^{ N-1}
\ee^{N\big\{-{2 \pi^2 \over A} +1 - \sum\limits_{k=1}^\infty
{2k \choose k}
{({A/ 4 \pi^2})^k\over
 k(k+1)}\big\} } ,
}
the effect of which is to shift the critical area to $A=\pi^2$.

Can  such a phase transition occur in four dimensional QCD?
Consider putting the theory in a box of volume $V$.
Then for small $V$ we can trust perturbation theory as well as the  dilute gas
treatment of instantons.
The finite size of the the box will cutoff the size of the instantons,
thus they will be governed by
weak coupling and have small density. For large $N$ they will therefore
disappear. However,
as the size of the  box increases the  relevant coupling will increase and
we could
induce just such a phase transition. This would be very similar in nature
to the case of QCD$_2$, which does undergo a phase transition for finite
volume as the volume is increased.

Similarly, one might contemplate a
phase transition
for physical observables. Consider the expectation value of a Wilson loop of
area $A$. As $A$
is decreased perturbation theory becomes more and more reliable.
The contribution of instantons will be cutoff since only instantons
of the size of
the loop and smaller will affect
the loop, thus again the instanton density will be governed by weak coupling.
As the size of the loop is increased the density increases.
It could be that for $N=\infty$ there
is again a phase transition. This would imply, among the rest,
that the string representation
of QCD, which is a $1/N$ representation, would break down for certain
questions, such as the free energy of QCD in a small box, or the
behavior of small Wilson loops. The string representation  of QCD would
then only be applicable to the infra-red domain.

To investigate this possibility let us write the instanton density in
the dilute gas approximation.
This calculation was done a long time ago \REFbernard. The instanton density is
given as,
\eqn\qcdinstanton{ \int {d \rho \over \rho^5} \thinspace
0.465994 N^3 \biggl[\left({0.432133 {\rm e}\over
\lambda(\rho)}\right)^2
{\rm e}^{-{1\over \lambda(\rho)} + O(\lambda)}\biggr]^N,
}
where the integral is over the scale size of the
instanton and
\eqn\coupling{ {1 \over \lambda(\rho)} =
{8\pi^2\over g^2(\rho)N}
={ 8\pi^2\over g^2N} -{ 11\over 3} \ln(  \Lambda_{\rm PV}\rho) .
}
This calculation contains the instanton action, the one-loop
determinant and the
volume of the collective modes. $ \Lambda_{\rm PV}$ is the renormalization
scale in the Pauli-Villars regulation scheme. The power of the inverse coupling
arises
from the fact that  the instanton
is embedded in an $SU(2)$ subgroup of $SU(N)$ and  that
there are $(N^2-1) -((N-2)^2 -1)-1=4N-5$ gauge zero modes
plus $4$ translational zero-modes and one scale zero mode.
In the ordinary vacuum there is no cutoff on the size
of instantons, $\lambda(\rho)$ grows for large
$\rho$ and this approximation breaks down in the infra-red.
In the situations we are imagining
the finite size of the box imposes a cutoff on the size of the
relevant instantons, which increases as the
size of the box, or the Wilson loop is increased.
For small boxes or loops the effective coupling will never be large,
and the instanton density will vanish as $N\to \infty$.
But as in QCD$_2$ we could  contemplate a phase
transition as the infrared cutoff is removed and the effective coupling grows.

The question of whether or not
instantons survive in the large $N$ limit was explored by
H. Neuberger \REFNeuberger
\foot{We thank Herbert Neuberger for bringing this
reference, which we had forgotten
in an earlier version of this paper,  to  our attention.}.
As he noted, in  order to have a large $N$
transition we require  that $c_2 c_3/c_1 \geq   {\rm e}$.
In four dimensional QCD
$$c_1=1, \ c_2=0.432133\thinspace {\rm e}, \ c_3=2 \ \ \ \Rightarrow
c_2 c_3/c_1= 0.864267\thinspace {\rm e} <  {\rm e}.$$
Therefore the above requirement is not satisfied and
the maximum instanton density
behaves as $(0.864267)^N \to 0$, as $N\to \infty$.  However it is known
that the instanton density depends on the method of regularization.
Different methods of regularization correspond in one loop order to a
shift in the renomalization scale \REFhas, \REFdash\ \
$\Lambda \to \Lambda'$, and thus to a multiplicative change  in the
density of the instantons
$\rho(\Lambda')={(\Lambda/ \Lambda')}^{11N/3}\rho(\Lambda)$.
Thus if we use ${\rm M{\bar S}}$ regularization
$(\Lambda_{\rm PV}/\Lambda_{\rm M{\bar S}})=
\exp(1/12)$ \ \REFdash\  and $c_2$ is enhanced by a factor of $1.16507$,
in which case the maximal instanton density behaves as
$( 1.00656)^N \to \infty $ as $N\to \infty$. Alternatively
if we use a momentum-space subtraction procedure (in Feynman gauge) then
$(\Lambda_{\rm PV}/\Lambda_{\bar{\rm MOM}})= 0.37594$,
$c_2$ is decreased by a factor of
$0.166$ and therefore
the maximal instanton density behaves as
$(0.14378)^N \to 0$ as $N\to \infty$. The regularization that makes the
instanton density largest is the lattice regularization, for which
$(\Lambda_{\rm PV}/\Lambda_{\bar{\rm L}})=42.299$.
In fact Neuberger argued that for this lattice regularization
instantons do survive as $N\to \infty$. His concern was mostly in the
contribution of instantons to the $\beta$-function, and the huge jump,
using  lattice $\Lambda$, suggested a change in physics at
the scale where instantons became relevant.

The issue of whether or not a large $N$
phase transition occurs for physical observables
as we vary some physical parameter
(the volume of space-time or the size of a Wilson loop) is
surely independent of the regularization scheme. Therefore
it is clear that higher order calculations of the instanton
density, which are also regularization scheme
dependent, must be performed to settle the issue.
To the extent
that we can trust the above, one-loop,
calculation of the instanton density, using the momentum-space subtraction
or Pauli-Villars regularization schemes, we would conclude  that
{\it there is no large $N$ instanton induced phase transition in QCD$_4$}.

The above considerations are consistent with the idea that
in QCD$_4$ there are no large $N$
phase transitions\foot{There might very well be large $N$ phase
transitions in lattice QCD as one
approaches the continuum limit, but these are lattice artifacts.}.
However due to the regularization dependence, and the absence of higher order
calculations, it  remains an open issue. There is another reason why one might
suspect that no large $N$ phase transition is likely as one varies the size of
a Wilson loop. More precisely one could have explored a phase transition in a
correlation function of local operators, say $<0| \bar \Psi \Psi(x)\thinspace
\bar \Psi \Psi(0)|0>$, as a function of $x$. For small $x$ instantons
are suppressed
and one could ask whether their contribution to this correlation function leads
to a phase transition as one increases $x$. This  cannot happen  if the large
$N$ theory
is well-behaved, since it would be inconsistent with  Euclidean analyticity of
the amplitudes. In the case of a Wilson loop the situation   is somewhat less
clear. We do not know of the existence of a
spectral representation of the loop which would imply
analyticity in the radius. However the analogy suggests
that a phase transition is unlikely.

If this is the case a QCD string theory is likely to be applicable
both for short and for long distances.

\newsec{The Double Scaling Limit.}

Let us now turn to the problem of the double scaling limit at the point
of the phase transition.
It is a typical situation in matrix models that when there is a phase
transition  all
coefficients of the $1/N$ expansion will be singular at $A=A_c$.
If the singularity has a special type
so that the power of singularity in $A-A_c$ grows linearly with the order of
expansion, $g$:
$$ N^{2-2g} f^{G=0}_{g}(A) \sim \big(N^2 (A-A_c)^{2-\gamma_{\rm str}}
\big)^{1-g},$$ then double scaling limit \REFgrmig\ can be defined by
$$ N\rightarrow \infty \, , \quad A \rightarrow A_c \, ; \quad
g_{\rm str}^{-2}\equiv N^2 (A-A_c)^{2-\gamma} \quad {\rm fixed}.$$
Such a limit is known to describe certain versions of two-dimensional
quantum gravity, the $1/N$ expansion becoming the expansion in powers
of the string coupling constant $g_{\rm str}$, the genus expansion of
string theory\foot {Note that the string theory we are contemplating
is not the string representation of QCD$_2$, for
which $g_{\rm str}=1/N$, but rather some kind of 2d gravity
we  obtain at the critical point $A=A_c$.}.

In the case of two-dimensional QCD
the triviality of the $1/N$ expansion in the weak coupling
phase prevents the direct identification of such limit. This might not be
the case if we approach the point of transition through the strong coupling
phase, $A>A_c$. In fact, even in the weak coupling phase there is
strong evidence that the double scaling limit exists. Indeed, the form of
singularity in the leading order in $N$, found by Douglas and Kazakov
\REFdouglas,
\eqn\sing{
N^2 f_{0}^{0}(A)_{strong} -N^2 f_{0}^{0}(A)_{weak} \sim N^2 (A-A_c)^3
}
implies that we need to fix
$$g_{\rm str}^{-2}=N^2(\pi^2 -A)^3, $$
that is, $\gamma_{\rm str}=-1$.

We can test this conjecture by looking closely at the free energy
\instantonenergy. If $A \rightarrow \pi^2$,
$$\gamma\Bigl({A\over \pi^2}\Bigr)\simeq {2\over 3}\Bigl(1- {A\over \pi^2}
\Bigr)^{3/2}, $$
and
$$ F(A, N) \sim {\rm regular\  terms} -{(-)^N\over \sqrt{8\pi N}}
\Bigl(1- {A\over \pi^2}\Bigr)^{-1/4} \ee^{-{4\over 3 \pi^3} (N^2 (\pi^2-A)^3)
^{1/2}}.$$
We see that $N$ in the exponential disappears so that it takes the form
\eqn\nonpt{
-{1\over 2\sqrt{2}} N^{-1/3} g_{\rm str}^{1/6}
{\rm exp}\Bigl(-{4\over 3 \pi^3 g_{\rm str}}\Bigr).
}
This is the correct behavior of the  nonperturbative terms that one expects.
Note the characteristic $\exp[-1/ g_{\rm str}]$. In fact,
\nonpt \ is the small $g_{\rm str}$ asymptotics of the complete nonperturbative
expression that will arise in the double scaling limit.

Let us now try to find this nonperturbative expression in full. To do this
we will need to know the behavior of the recursion coefficients $R_n$
in the immediate vicinity of the critical point $A_c=\pi^2$. In the following
it will be convenient to introduce the {\it scaling variable}
\eqn\scaling{x\equiv n_{\rm cr}^{2/3}\Bigl(1-{n\over n_{\rm cr}}\Bigr).
}
Keeping $g_{\rm str}$ fixed is tantamount to fixing $x$, and taking
$N \rightarrow \infty$ is the same as
considering $n_{\rm cr}\rightarrow \infty$\
(since $n_{\rm cr}=\pi^2 N/A$, $A$ being finite).

We can extract some information about $R_n$ from our asymptotic formula
\rass :
\eqn\dbs{\eqalign{
R_n\simeq& {n_{\rm cr}^2\over \pi^2}+ (-)^n
\sqrt{2 \over \pi^5} n_{\rm cr}^{5/3}
x^{-1/4} \ee^{-{4\over 3} x^{3/2}}
+{\cal O}\big(n_{\rm cr}^{4/3}\big),\cr
&\quad n_{\rm cr}\rightarrow\infty, \quad x \ {\rm fixed}.\cr
}}
We have taken into account that
$${n\over n_{\rm cr}}=1- {x\over n_{\rm cr}^{2/3}}\rightarrow 1$$
in the double scaling limit\foot{Again, this gives large $n_{\rm cr}$, and
{\it large} $x$ asymptotics of $R_n$, like \nonpt.}. Since
the correction to $R_n$ has an
alternating sign, it is natural to expect this correction to
behave differently for odd and even $n$.

To see how $R_n$ behaves at finite $x$, we consider the following ansatz:
\eqn\ans{
R_n={n_{\rm cr}^2\over \pi^2} +
n_{\rm cr}^{5/3}f_{1\pm}(x)+n_{\rm cr}^{4/3}f_{2\pm}(x)
+n_{\rm cr} f_{3\pm}(x)+ {\cal O}\big(n_{\rm cr}^{2/3}\big), \quad n_{\rm cr}
\rightarrow\infty.
}
The subscripts $\pm$ refer to the even and odd $n$, respectively.

We will now write down a system of ordinary differential equations
for the functions $f_{1\pm}(x), f_{2\pm}(x), f_{3\pm}(x)$. This system follows
from the Volterra equation \volterra.
Replacing $\alpha=A/2N = \pi^2/2n_{\rm cr}$
we can rewrite \volterra\ as
\eqn\vol{
R_{n+1}-R_{n-1}= {2 n_{\rm cr}^2\over \pi^2} {d\over d n_{\rm cr}}\ln R_n.
}
Then, considering that
$$x_{n\pm1}=n_{\rm cr}^{2/3} \Bigl(1- {n\pm1 \over n_{\rm cr}}\Bigr)=
x_n\pm {1\over n_{\rm cr}^{1/3}} ,
$$
we obtain, say, for odd $n$,
\gdef\fp#1{f_{#1+}\Bigl(x-{1\over n_{\rm cr}^{1/3}}\Bigr)}
\gdef\fm#1{f_{#1+}\Bigl(x+{1\over n_{\rm cr}^{1/3}}\Bigr)}
\gdef\f#1{\biggl(\fp#1 - \fm#1\biggr)}
\eqn\difference{\eqalign{
R_{n+1}-R_{n-1}=n_{\rm cr}^{5\over 3}&\f1 \cr
+n_{\rm cr}^{4\over 3}&\f2 \cr
+n_{\rm cr}&\f3 +\ldots\cr
= -2 n_{\rm cr}^{4/3}f_{1+}^{\prime}&(x)-2 n_{\rm cr}f_{2+}^{\prime}(x)
-2 n_{\rm cr}^{2/3} \big(f_{3+}^{\prime}(x) +
{1\over 6}f_{1+}^{\prime\prime\prime}(x)\big)+\ldots.\cr
}}

On the other hand, to calculate the right hand side of \vol,
${2n_{\rm cr}^2\over \pi^2}{d\over d n_{\rm cr}}\ln R_n$, we must remember that
$R_n$ depend on $n_{\rm cr}$ both explicitly and implicitly, through $x$,
$${\partial x\over \partial n_{\rm cr}}\biggl|_{n=const}
= {1\over n_{\rm cr}^{1/3}}-{1\over 3}{x\over n_{\rm cr}}.$$

The final result of this computation is
\eqn\derivative{
\eqalign{
-{d \over d\alpha}\ln R_n(\alpha)&={2 n_{cr}^2\over \pi^2}{d \over d n_{cr}}
\ln R_n(\alpha)= {4 n_{cr}\over \pi^2}
 +2 n_{cr}^{4\over 3} f_{1-}^{\prime}(x)\cr
&+2 n_{cr}\big(f_{2-}^{\prime}(x)- \pi^2 f_{1-}(x) f_{1-}^{\prime}(x)\big)\cr
&+2 n_{cr}^{2\over 3} \Bigl(f_{3-}^{\prime}(x)- {1\over 3}\big( f_{1-}(x)
+ x f_{1-}^{\prime}(x)\big) \cr
&\quad -\pi^2\big(f_{1-}(x) f_{2-}^{\prime}(x)+
f_{2-}(x) f_{1-}^{\prime}(x)\big) +\pi^4 f_{1-}^2 (x)f_{1-}^{\prime}(x)\Bigr)
\cr
&+\ldots
.\cr
}}
Equating this to \difference\ we get
\eqn\system{\eqalign{
&f_{1+}^{\prime}(x)+f_{1-}^{\prime}(x)=0, \cr
&f_{2+}^{\prime}(x)+f_{2-}^{\prime}(x)+{2\over \pi^2}=
\pi^2 f_{1-}(x)f_{1-}^{\prime}(x), \cr
&\eqalign{f_{3+}^{\prime}(x)+f_{3-}^{\prime}(x)=& {1\over 3}\big( f_{1-}(x)
+ x f_{1-}^{\prime}(x)\big)+\pi^2\big(f_{1-}(x) f_{2-}^{\prime}(x)+
f_{2-}(x) f_{1-}^{\prime}(x)\big) \cr
&-\pi^4 f_{1-}^2 (x)f_{1-}^{\prime}(x)
-{1\over 6} f_{1+}^{\prime\prime\prime}(x).\cr}\cr
}}

Similarly, writing down the Volterra equation for even $n$,
it is easy to see that
the same set of equations will hold if we replace all $f_{j-}(x)$ by
$f_{j+}(x)$, and vice versa. It follows then from the third equation \system\
that
\eqn\third{
\eqalign{f_{3+}^{\prime}(x)+f_{3-}^{\prime}(x)=& {1\over 3}\big( f_{1-}(x)
+ x f_{1-}^{\prime}(x)\big)+\pi^2\big(f_{1-}(x) f_{2-}^{\prime}(x)+
f_{2-}(x) f_{1-}^{\prime}(x)\big) \cr
&-\pi^4 f_{1-}^2 (x)f_{1-}^{\prime}(x)
-{1\over 6} f_{1+}^{\prime\prime\prime}(x)\cr
=& {1\over 3}\big( f_{1+}(x)
+ x f_{1+}^{\prime}(x)\big)+\pi^2\big(f_{1+}(x) f_{2+}^{\prime}(x)+
f_{2+}(x) f_{1+}^{\prime}(x)\big) \cr
&-\pi^4 f_{1+}^2 (x)f_{1+}^{\prime}(x)
-{1\over 6} f_{1-}^{\prime\prime\prime}(x)\cr
}}
which, together with\foot{The first equation in \system\ can
be integrated to get $f_{1+}(x)=-f_{1-}(x)$, since both $f_{1+}(x)$
and $f_{1-}(x)$ decay exponentially at large $x$, see \dbs.}
$f_{1+}(x)=-f_{1-}(x)$, implies
\eqn\onemore{
{2\over 3}(f_{1-}+xf^{\prime}_{1-}) + {1\over 3}f_{1-}^{\prime
\prime\prime}-2 \pi^4 f_{1-}^2 f_{1-}^{\prime}+
\pi^2 \big(f_{1-}(f_{2+}+f_{2-})\big)^{\prime}=0.
}
Integrating the second equation in \system, we derive
\eqn\fsecond{
f_{2+}+f_{2-}= -{2\over \pi^2} x+{\pi^2\over 2}f_{1-}^2 + C
}
where $C$ is the constant of integration. Substituting this into \onemore\
and integrating once again, we get an equation for $f_1(x)\equiv f_{1-}(x)$:
\eqn\ffirsteq{
f_1^{\prime \prime}-4x f_1 -{\pi^2\over 2}f_1^3= 3D-3\pi^2 C f_1.
}
The constants $C$ and $D$ can be fixed using the large $x$ asymptotics
\dbs:
\eqn\assy{
f_1(x)\simeq \sqrt{{2\over \pi^5}}x^{-{1\over 4}}\ee^{-{4\over 3}x^{3/2}},
\quad \ x\rightarrow +\infty
}
to be $C=D=0$.

Therefore, $f_1(x)$ obeys the Painleve
\uppercase\expandafter{\romannumeral 2}\  equation
\eqn\pain{
f_1^{\prime\prime}-4 x f_1 -{\pi^2\over 2}f_1^3=0.
}

Now we are in a position to evaluate the free energy in the double scaling
limit. However, it is more convenient to consider its second derivative with
respect to the area (the ``specific heat capacity") $d^2 F_N(A)/dA^2$.
Since
$$ F_N(A)= N\ln h_0\Bigl(\alpha= {A\over 2N}\Bigr)+
\sum_{n=1}^{N-1}(N-n) \ln R_n(\alpha), $$
it is easy to see that, as a consequence of \rel, \volterra,
$${dF_N(A)\over dA}=-{1\over N}\Bigl[{1\over 2}R_N + \sum_{n=1}^{N-1}R_n\Bigr]
$$
and
\eqn\capacity{
{d^2 F_N(A)\over dA^2} ={1\over 4N^2}R_N (R_{N+1}+R_{N-1}).
}
With the aid of our double scaling ansatz \ans\ and using \fsecond\ we get
$$\eqalign{
{d^2 F_N(A)\over dA^2} = &{n_{cr}^4\over 2\pi^4 N^2}\biggl[1+ {\pi^2\over
n_{cr}^{2/3}}\big(f_{2-}(x)+f_{2+}(x)- \pi^2 f_{1-}^2(x)\big)
+{\cal O}\Bigl({1\over n_{cr}}\Bigr)\biggr]\cr
= &{n_{cr}^4\over 2\pi^4 N^2}\biggl[1-2\Bigl(1-{n\over n_{cr}}\Bigr)-
{\pi^4\over 2 n_{cr}^{2/3}} f_1^2(x) + {\cal O}\Bigl({1\over n_{cr}}
\Bigr)\biggr]
.\cr}
$$
Since at $A=A_c$ one has $n_{cr}=N$, we see that
the part of the specific heat,
singular at the transition point, is given by
\eqn\sheatsing{
\biggl({d^2 F_N(A)\over dA^2}\biggr)_{sing}=-{n_{cr}^{4/3}\over 4}f_1^2(x)
+\ldots.
}

The string equation \pain\ and the shape of the specific heat \sheatsing\
we have obtained are identical to the ones encountered in the double
scaling limit of the unitary one-matrix model. This suggests that the
theories defined by the two-dimensional QCD at the transition point
and by the one-plaquette model \REFgrwit, \REFperiwal\ ({\it i.e.}
one-matrix model) of QCD
may belong to the same universality class. In fact it is possible to
construct a very special one-plaquette model, adjusting its potential
to make it {\it exactly equivalent\/} to the two-dimensional QCD on a sphere.
We will discuss this point elsewhere.

\newsec{Conclusions}

We have investigated the $1/N$ expansion of the weak
coupling phase of two-dimensional QCD on a sphere. We demonstrated  that the
transition from the weak to the   strong coupling phase is induced by
instantons.
We   found a double scaling limit of the theory at the point of the
phase transition,
and determined the value of string susceptibility to be $\gamma_{str}=-1$.

We also explored the possibility of instanton induced large $N$
phase transitions in four dimensional QCD and presented a
calculation that shows that these are unlikely. There
are, of course, other possible sources of large $N$ phase
transitions, but our result lends support to the possibility that a
string representation of QCD$_4$ could be valid at all distances.
This  leaves us with the conceptual problem to understand
how strings manage to
behave as particles at short distances? On the other hand it
offers us the hope of being able to
test a string representation of QCD$_4$ by comparing it with the known,
perturbatively accessible, short distance structure of the theory.

\listrefs

\end